\documentclass[aps,prl,twocolumn,longbibliography,superscriptaddress,floatfix]{revtex4-1}
\usepackage{amsfonts}
\usepackage{mathtools}
\usepackage{graphicx}
\usepackage{epsfig}
\usepackage{dcolumn}
\usepackage{bm}
\usepackage{amsmath}
\usepackage{graphicx}
\usepackage[latin1]{inputenc}
\usepackage{ulem}
\usepackage{epstopdf}
\usepackage{subfigure}
\usepackage{color}
\usepackage{amsthm}
\usepackage{newlfont}
\usepackage{graphicx}
\usepackage{epstopdf}
\usepackage{appendix}
\usepackage[breaklinks=true]{hyperref}
\usepackage{breakcites}
\usepackage{textcomp}
\usepackage{appendix}
\usepackage{multirow}	
\usepackage{color}
\usepackage{amssymb}
\usepackage{epsfig}
\usepackage{bm}
\usepackage[american]{babel}
\usepackage{braket}
\hypersetup{colorlinks=true,linkcolor=blue,citecolor=blue,filecolor=blue,urlcolor=blue,pdfstartview=FitH}

\begin{document}
\title{
Fragile Many Body Ergodicity
}
\author{Thudiyangal Mithun}
\affiliation{Department of Mathematics and Statistics, University of Massachusetts, Amherst Massachusetts 01003-4515, USA}
 \affiliation{Center for Theoretical Physics of Complex Systems, Institute for Basic Science, Daejeon 34051, Korea}
 \author{Carlo Danieli}
\affiliation{Max Planck Institute for the Physics of Complex Systems, N\"othnitzer Str. 38, 01187 Dresden, Germany}
 \affiliation{Center for Theoretical Physics of Complex Systems, Institute for Basic Science, Daejeon 34051, Korea}
  \author{M. V. Fistul}
\affiliation{Center for Theoretical Physics of Complex Systems, Institute for Basic Science, Daejeon 34051, Korea}
\affiliation{Theoretische Physik III, Ruhr-Universit\"at Bochum, Bochum 44801, Germany}
\affiliation{Russian Quantum Center, National University of Science and Technology "MISIS", 119049 Moscow, Russia}
\author{B. L. Altshuler}
 \affiliation{Physics Department, Columbia University, New York 10027, USA}
  \affiliation{Center for Theoretical Physics of Complex Systems, Institute for Basic Science, Daejeon 34051, Korea}
 \author{Sergej Flach}
 \affiliation{Center for Theoretical Physics of Complex Systems, Institute for Basic Science, Daejeon 34051, Korea}
 \affiliation{Basic Science Program(IBS School), Korea University of Science and Technology(UST), Daejeon 34113, Korea}

\begin{abstract}
Weakly nonintegrable many-body systems can restore ergodicity 
in distinctive ways depending on the range
of the interaction network in action space. Action resonances seed chaotic dynamics into the networks. Long range networks provide well connected resonances with ergodization controlled
by the individual resonance chaos time scales.
Short range networks instead yield a dramatic slowing down of ergodization in action space,
and lead to rare resonance diffusion. We use Josephson junction chains as a paradigmatic study case. We exploit finite time average distributions to characterize the thermalizing dynamics of actions. We identify a novel action resonance diffusion regime responsible for the slowing down. We extract the diffusion coefficient of that slow process and measure its dependence on the proximity to the integrable limit. Independent measures of correlation functions confirm our findings. The
observed fragile diffusion is relying on weakly chaotic dynamics in spatially isolated action resonances. It can be suppressed, and ergodization delayed, by adding weak action noise, as a proof of concept.
\end{abstract}
\maketitle

The conventional perception of evolving dynamical systems with a macroscopic number of degrees of freedom (DoF) is them being in a state of thermal equilibrium, i.e. ergodic. This assumes all allowed microstates having the same probability. It goes along with trajectories visiting the vicinity of all points of the available phase space (i.e. the phase space subject to constraints due to integrals of motion such as e.g. the energy), and infinite time averages equaling available phase space averages \cite{book:huang87}. Statistical physics approaches were paved by Gibbs and Boltzmann and provide a straight connection between microcanonical dynamics and the emergence of canonical distributions \cite{book:huang87}. The more interesting it is to study 
cases when this connection is not evident, eroding, or even missing. This can happen (i) for dynamics in the proximity of an integrable limit,
(ii) for dynamics in the proximity of nonergodic sets of measure zero (such as periodic orbits), and (iii) for dynamics driven out of ergodicity
due to additional constraints (e.g. condensation). Fermi-Pasta-Ulam-Tsingou problems 
\cite{FORD1992FPU,fpugen,FlachIvanchenkoKanakovPRL2005,gallavotti2007,Danieli:2017} can be associated with (ii), and non-Gibbs states for interacting Bose 
lattice gases \cite{rasmussen00,Iubini:2013,Mithun:2018,ChernyEnglFlach19,Iubini:2019dynamical} with (iii). 
As for (i), 
the celebrated Kolmogorov-Arnold-Moser (KAM) theorem is available \cite{Moser2001}, but applies
to systems with finite numbers of DoF and dictates weakly non-integrable dynamics to be non-ergodic on a finite measure set of invariant tori, while being ergodic on the complementary one (Arnold diffusion \cite{DiacuHolmes1999}). The KAM borders are assumed to quickly diminish with increasing DoF numbers \cite{Moser2001}. What lies beyond those borders for macroscopic systems? The expectation that Gibbs and Boltzmann take over, was shattered by recent results on Many-Body Localization (MBL) \cite{Basko2006metal,NandkishoreHuse2015} which show that certain quantum many-body systems can resist thermalization at finite distance from integrable limits. 
With most analytical results being non-rigorous, and computations notoriously heavy due to exploding Hilbert space dimensions, 
the weakly touched field of ergodization and thermalization of corresponding classical many body systems is in the focus of this work.

Networks of weakly coupled superconducting grains are one of the few paradigmatic examples of systems where the above scenaria
have been considered \cite{Escande:1994,Pino:2016,Mithun:2019}. Also, related networks of interacting anharmonic oscillators were used to argue for and show
the existence of two different classes of nonintegrable perturbations of an integrable Hamiltonian $H_0(\{J_k\})$, with a countable
set of actions $J_k$ ($k$ being an integer) \cite{Danieli2019dynamical}. Nonintegrable perturbations $H_1(\{J_k,\Theta_k\})$ typically span long range or short range networks (LRN or SRN) in the action-angle space (here $\Theta_k$ are the canonically conjugated angles). 
A reference action $J_k$ in that network is coupled to 
$R_k \times L_k$-tuples of other action-angle pairs. $L_k$ is typically a single digit integer.
$R_k$ however can be intensive (SRN) or extensive (LRN) \cite{Danieli2019dynamical}.

Consider a typical LRNs, and translationally invariant two-body interactions $L_k=3$, $R_k \sim N^2$ with $N$ being the volume (system size) \cite{Danieli2019dynamical}.
Chaotic dynamics can develop in a given $L$-tuple on a time scale $T_{\Lambda}=1/\Lambda$ where $\Lambda$ is the typical
(largest) Lyapunov exponent in the system. Chaotic dynamics develops due to nonlinear resonances which take place
when ratios of network matrix elements to certain frequency differences are large \cite{Chirikov:1979}.
In the proximity to an integrable limit the network matrix elements scale with $1/N$ \cite{Danieli2019dynamical}.
Therefore the probability
for an $L$-tuple to be resonant will be $ \pi_r/N \ll1$ where $\pi_r \ll 1$ is an intensive measure of the distance to the integrable limit. However, there are $R_k \sim N^2$ $L$-tuples in which one given action is involved. Therefore the probability
$\Pi_k$ that the reference action is in resonance with at least one of its $L$-tuples (i.e. satisfying the resonance condition)
turns exponentially close to unity: 
$\Pi_k  \approx 1- (1-\pi_r/N)^{R_k} \approx 1- e^{-R_k \pi_r/N}$ for such a macroscopic LRN. It follows that LRNs  thermalize homogeneously in action space,
see e.g. \cite{Danieli2019dynamical}.

On the contrary, SRNs show anomalously slow ergodization and thermalization dynamics in proximity to an integrable limit \cite{Mithun:2019,Danieli2019dynamical}.
Since $R_k$ is now intensive and $N$-independent, the resonance probability $\Pi_k \sim \pi_r$ is small, resonances are rare, and thermalization is delayed until
resonances were able to migrate through the whole system. Thermalization is expected to be a highly inhomogeneous
process in action space.

In this work we quantitatively describe the dynamics of thermalization by making use of 
finite time average (FTA) distributions. We
(a) observe a novel regime of action diffusion, and
extract the diffusion coefficient as a function of the proximity to the integrable limit,
(b) show the connection between the dynamics of FTA distributions and auto-correlation functions and predict and observe
algebraic decay of correlations in time, and
(c) finally predict the
diffusion delay through action noise destroying resonances and provide computational evidence of the delay.

We consider the Hamiltonian 
\begin{equation}
\small  H(q,p) =\sum_{n=1}^{N}\bigg[\frac{p_n^2}{2}+E_J(1-\cos(q_{n+1}-q_n)) \bigg], 
\label{eq:rotor1}
\end{equation}
describing the dynamics of a chain of $N$ superconducting islands with nearest neighbor Josephson coupling in its classical limit.
This model is equivalent to a 1D XY chain or simply a coupled rotor chain with rotor momenta $p_n$ and angles $q_n$.
$E_J$ controls the strength of Josephson coupling and will be compared to the energy density $h=H/N$.
The equations of motion of Eq.~(\ref{eq:rotor1}) read
\begin{equation}
\dot{q}_n =  p_n\;,\;\; \dot{p}_n   
= E_J \big[  \sin(q_{n+1}-q_n)+\sin(q_{n-1} - q_n)  \big].
\label{eq:rotor_eqm}
\end{equation}
We apply periodic boundary conditions $p_1 = p_{N+1}$ and $q_{1} = q_{N+1}$.
The system has two conserved quantities: the total energy $H$ and the total angular momentum $P=\sum_{n=1}^N p_n$. 
We will choose $P=0$ without loss of generality. 
A SRN limit is obtained for $E_J/h \rightarrow 0$, with
$H_0=\sum_{n=1}^{N}\frac{p_n^2}{2}$ and $H_1=\sum_{n=1}^{N} E_J(1-\cos(q_{n+1}-q_n))$.
The actions $J_n \equiv p_n$, and the angles $\Theta_l  \equiv q_n$.
Note that the opposite limit $E_J/h \rightarrow \infty$ (not further studied in this work) yields a LRN network 
as discussed in the introduction.

The microcanonical dynamics of (\ref{eq:rotor1}) explores the available phase space $\Gamma$.
The phase space average of an observable $f(\vec{X})=\langle f \rangle \equiv \frac{1}{Z} \int f(\vec{X}) d\Gamma \;,\; Z = \int d\Gamma$. Here 
$\vec{X}$ is a point in $\Gamma$.
The ergodicity property is tested quantitatively by  showing that
the infinite time average of any observable $f(\vec{X})$ 
will be equal to its phase space average $\langle f \rangle $.
\begin{figure} [!htbp] 
            \includegraphics[scale=1.0,width=0.48\textwidth]{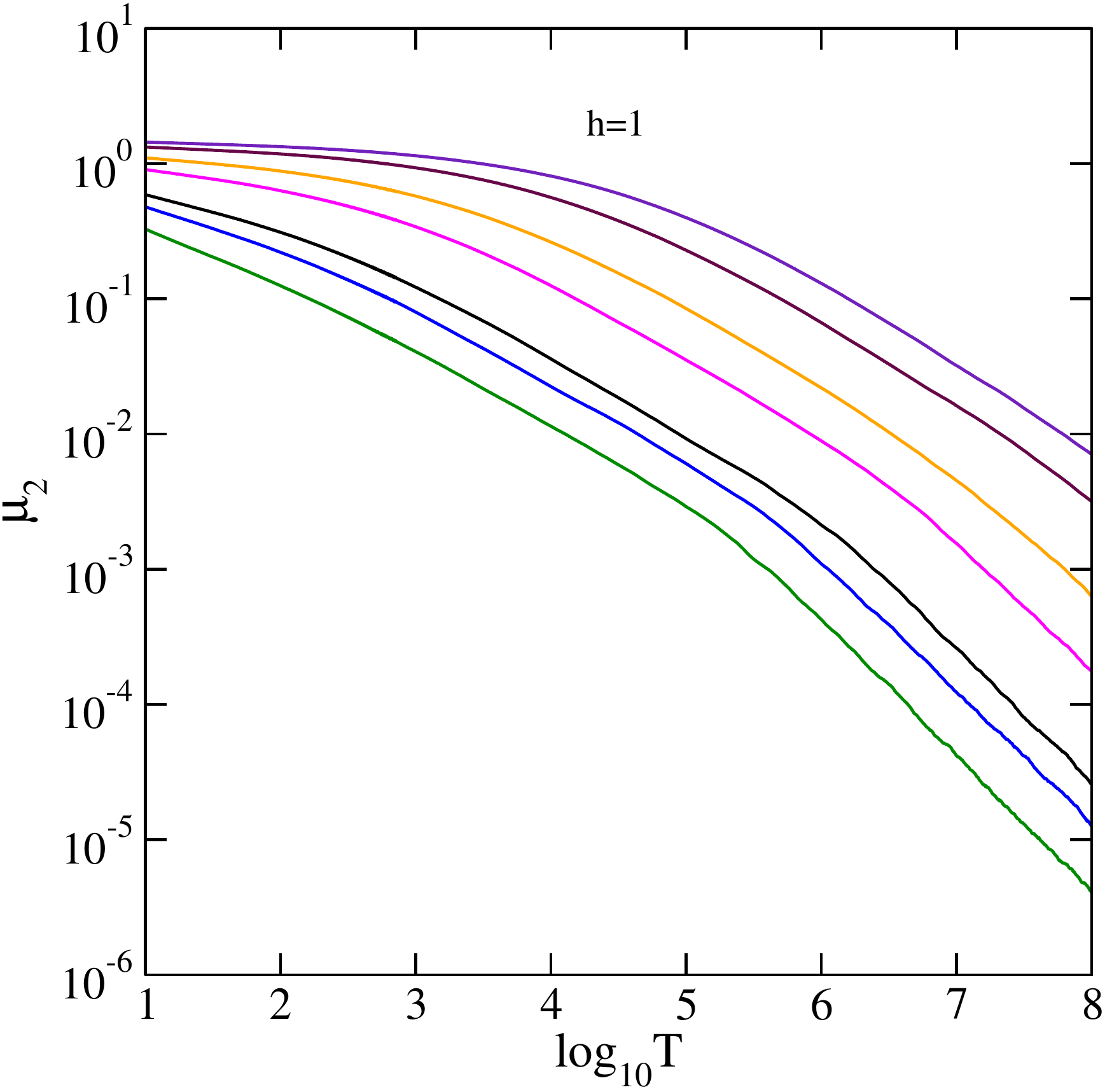}\\
  \caption{\label{Fig. m2t_all-n}{\footnotesize (Color online)  The time dependence of the second moment $\mu_{2}(T)$ for various values of $E_J$: $E_J=0.25, 0.3, 0.4, 0.5, 0.7, 0.8, 1.0$ from
top to bottom. Here $h=1$, $N=1024$ and $R=192$.
  }}
       \end{figure}
 \begin{figure} [!htbp] 
           \includegraphics[scale=1.0,width=0.48\textwidth]{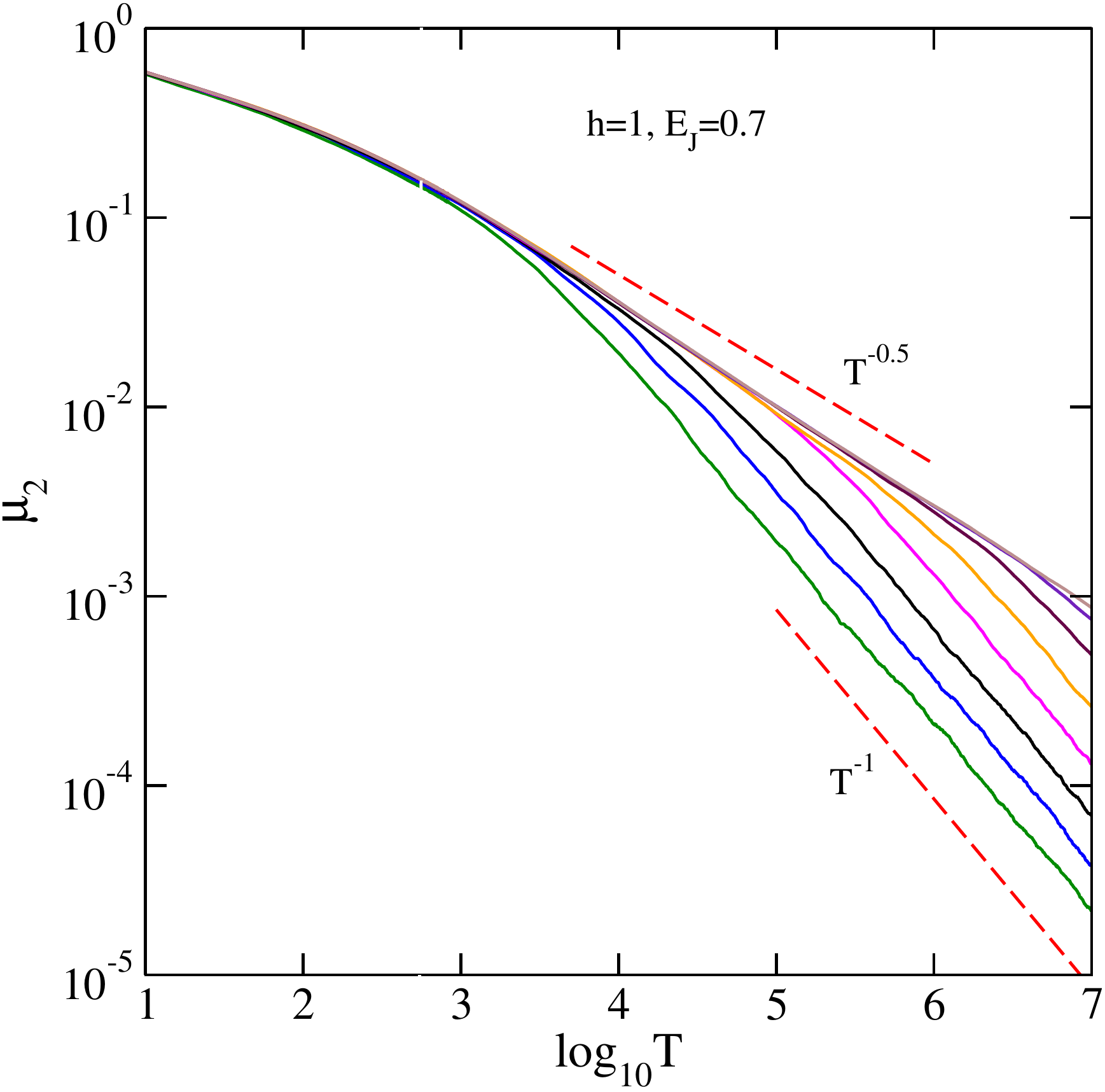}\\
  \caption{\label{Fig. m2p}{\footnotesize (Color online) The time dependence of the second moment $\mu_{2}(T)$ for different system sizes $N$: $N=2^6,2^7,2^8,2^9,2^{10},2^{11},2^{12},2^{13}$ from bottom to top. Here, $h=1$,  $E_J=0.7$, and $R=192$ .
  }}
     \end{figure} 
Lacking infinite times, we rather compute finite time averages, which depend on both the averaging time $T$, and
the initial condition $\vec{X}_0$:
\begin{equation}
f_T(\vec{X}_0)= \frac{1}{T}\int_{0}^{T}  f(\vec{X}(t)) dt \;,\; \vec{X}(t=0) = \vec{X}_0\;. 
\label{eq:ws_num}
\end{equation} 
For an ergodic system it follows
\begin{equation}
\lim_{T\rightarrow\infty} 
f_T(\vec{X}_0)  =   \langle f \rangle \; ,
\label{eq:mg}
\end{equation}  
for any choice of $\vec{X}_0$ except for a subset of measure zero.
Dense scanning of all initial points $\vec{X}_0$ over $\Gamma$ yields the {\sl finite time average distribution} 
$\rho(f;T)$ of the finite time averages $f_T(\vec{X}_0)$. It is a function of $f$, parametrically depends on $T$,
and is characterized by its moments
\begin{equation}
\mu_m(T) = \int f^m \rho(f;T) df \;,\; \mu_0=1\; .
\label{eq:rho_moments}
\end{equation}
It follows that the first moment $\mu_1 \equiv \langle f \rangle$ is invariant under variation of the averging time $T$. 
All higher moments will in general depend on $T$. For an ergodic system it follows
\begin{equation}
\lim_{T \rightarrow \infty} \mu_{m}(T) \rightarrow 0 \;, \; m \geq 2 \;.
\label{limitmoments}
\end{equation}
In our studies $\rho$ is close to a Gaussian distribution, which allows us to focus on $\mu_2$.

We use the actions (momenta) $p_n$ as the relevant slow observables
in the SRN proximity to the integrable limit $E_J \rightarrow 0$: $f \equiv p_n$. With $P=0$ it follows $\mu_1=0$.
The second moment $\mu_2(T)$ is then simply the variance of $\rho$, and further related to the momentum-momentum auto-correlation function 
$\mathcal{R}(t) = \lim_{\tau \rightarrow \infty} \frac{1}{\tau} \int_0^{\tau} p_l(\tau)p_l(t+\tau) d\tau$
as
$
\mu_2(T) = \frac{1}{T} \int_0^T \mathcal{R}(t) dt
$.
Under the usual assumption that the correlation function will have an exponential decay at large enough times (with ways to even
weaken the requirement) the conclusion is that $\mu_2(T \rightarrow \infty) \sim 1/T$.

The details of the integration methods are outlined in Ref. \cite{Mithun:2019}. We numerically integrate 
$R$ trajectories using symplectic integrators \cite{danieli2019computational}, where each initial point
was chosen by setting $q_n=0$, drawing $p_n$ from a Maxwell distribution, constraining $P=0$, rescaling all momenta such
that the desired energy density $h$ is obtained, and giving each trajectory a prethermalization run of $t_{prethermal}=10^6$. 
Since all actions $p_n$ are statistically equivalent, we measure them all and add all data into one pool which is used to compute
$\rho$. 
Fig.~\ref{Fig. m2t_all-n} shows $\mu_{2}(T)$ for $h=1$, $N=1024$ and $0.25 \leq E_J \leq 1$.
The variance $\mu_2(T)$ resists decay up to some characteristic ergodization time scale $T_E$, after which it turns decaying 
as expected, signalling restoration of ergodicity. Note that this time scale $T_E$ was assessed in Ref. \cite{Mithun:2019} and
is an intensive time scale.

\begin{figure} [!htbp] 
            \includegraphics[scale=1.0,width=0.44\textwidth]{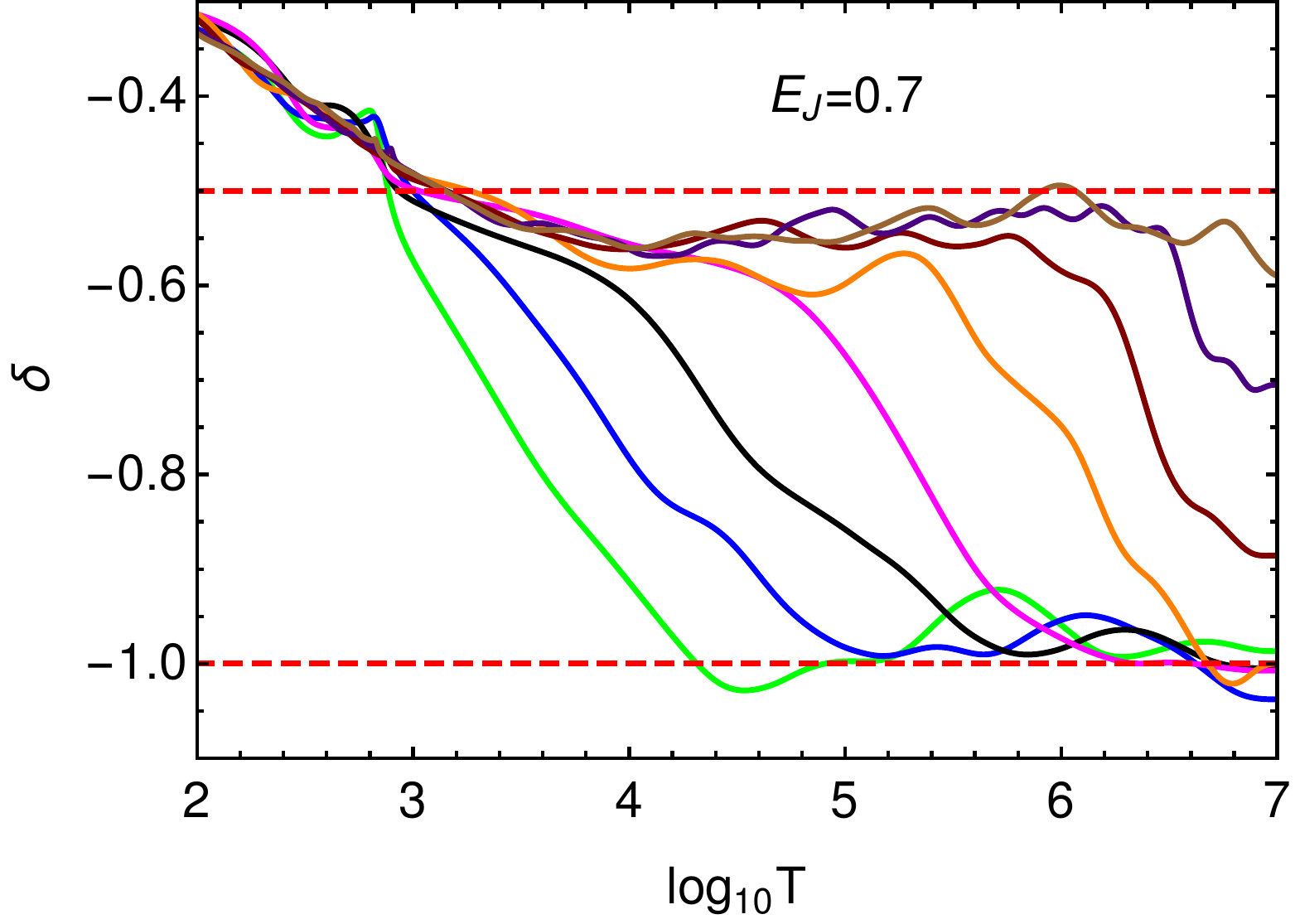}
  \caption{\label{Fig. m2ps}{\footnotesize (Color online)  The time dependence of the parameter $\delta=d(log_{10}\mu_2)/d(log_{10}T)$ for the curves in Fig.\ref{Fig. m2p}.
System size increases at $\delta=-0.8$ from left to right. The red dashed lines represent $\delta = -0.5$ and $\delta=-1$.
  }}
     \end{figure}

A close inspection of the size dependence of $\mu_2(T)$ is shown in 
Fig.~\ref{Fig. m2p} for $h=1$, $E_J=0.7$ and a variety of system sizes. 
We find that $\mu_2(T_E \leq T \leq T_D)$ loosely follows a $1/\sqrt{T}$ diffusive decay, which is followed by the anticipated
$1/T$ decay for $T_D \leq T$. The new time scale $T_D(N)$ is evidently system-size dependent. To support that finding, we
plot $\delta=d(log_{10}\mu_2)/d(log_{10}T)$ versus $T$ in Fig.~\ref{Fig. m2ps}. 
The curves clearly show intermediate saturation on a plateau with $\delta \approx -0.5$, and a subsequent decay at $T_D(N)$ down
to $\delta = -1$.
Since $T_D$ is increasing with system size,
we conjecture that $T_D(N \rightarrow \infty) \rightarrow \infty$, extending the $1/\sqrt{T}$ decay in $\mu_2(T)$ to infinite times for infinite size.
In turn this implies that the correlation function $\mathcal{R}(t) \sim 1/\sqrt{t}$ without any exponential cutoff in the same limit.
     
Let us discuss possible mechanisms leading to the observed behavior of $\mu_2(T)$ for different system sizes.   
We consider the occurence of a chaotic resonance to take place when first order perturbation theory for the evolution of a given
rotor at site $n$ breaks down. A simple calculation yields $\Delta_n^+ <E_J$ and $\Delta_{n}^-<E_J$ with $\Delta_n^\pm = |p_n(p_n-p_{n\pm 1})|$.
The presence of such resonantly coupled triplets of grains along the network generate chaotic dynamics 
and results in a Lyapunov exponent whose inverse yields a time scale $T_{\Lambda} \leq 10$ on the studied interval $0.25 \leq E_J \leq 1$ \cite{Mithun:2019}. 
It follows that $T_{\Lambda} \ll T_E,T_D$. 
The resonance probability can be easily computed as $\pi_r \sim (E_J/h)^2$. At variance to the LRN cases, the
SRN resonances are rare and inhomogeneously distributed over the system at any time. The
typical distance between consecutive chaotic triplets $l_r \sim (h/E_J)^2$  grows with reducing $E_J$ turning the resonances more sparse and rare \cite{Basko_2012}. 
The assumption of partial thermalization of actions involved in the rare resonances will still not lead to any
substantial observation of the onset of ergodization, simply because resonances are rare and separated by non-chaotic (regular) regions. 
To onset ergodicity instead, chaotic resonances have to diffusively migrate throughout the entire system \cite{Mithun:2019}. 
This presumably happens due to an incoherent detuning of the momenta of rotors 
in a neighbourhood of a given resonance. Once such a neighbouring rotor is sufficiently detuned, it could become resonant with its own neighbourhood forming a new resonance. 

\begin{figure} [!htbp] 
            \includegraphics[scale=1.0,width=0.48\textwidth]{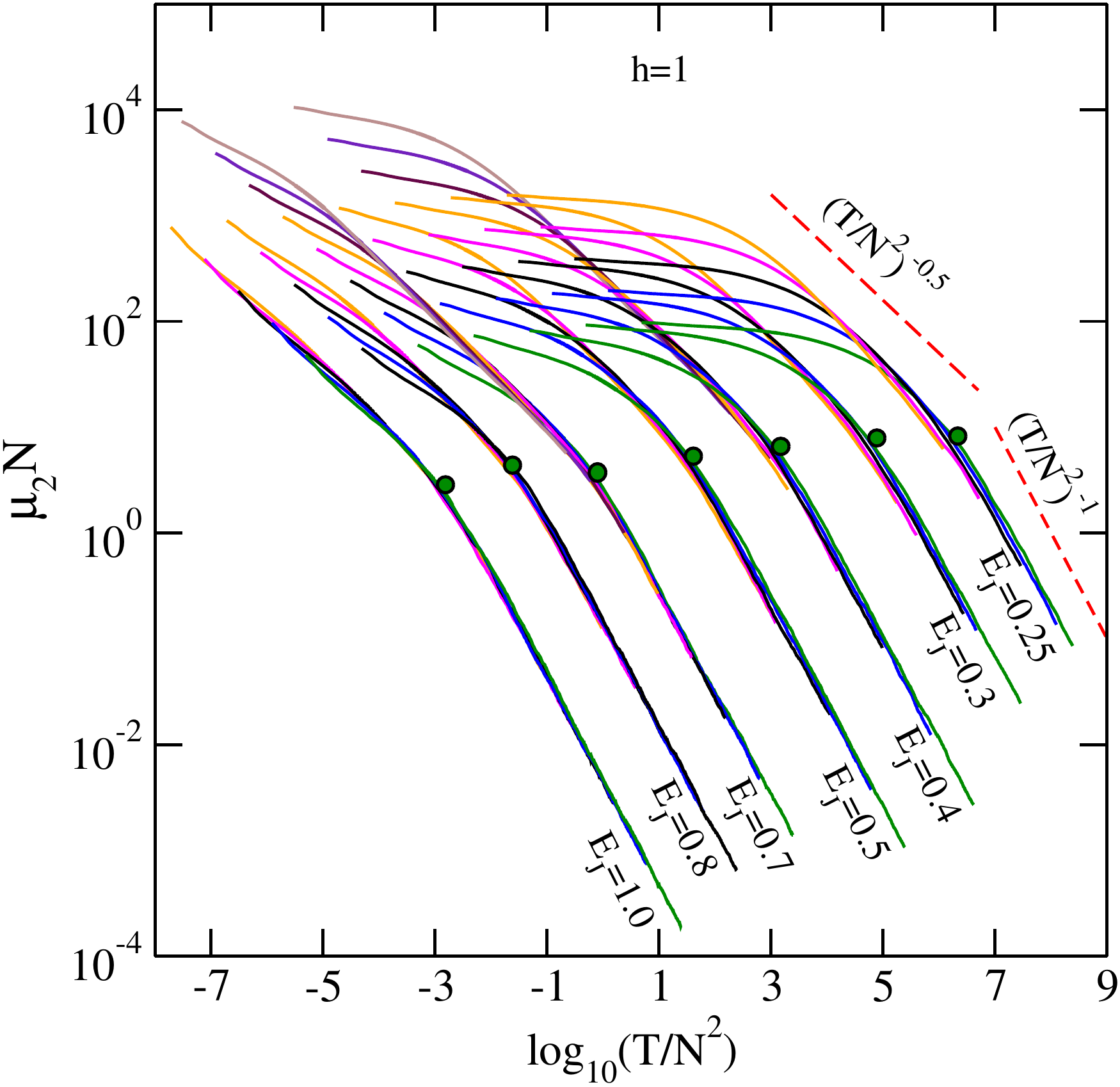}\\
  \caption{\label{Fig. Diffu_all-n}{\footnotesize (Color online) The rescaled time dependence of $\mu_2 N$ versus $b T/N^2$ for various $E_J$: $E_J=1.0,0.8,0.7,0.5,0.4,0.3,0.25$ from left to right.  The corresponding shift factor  $b=0.01,~0.1,~1,~10,~100,~1000,~10000$ 
is introduced for better visibility of the curves. Here $h=1$.
  }}
     \end{figure}

\begin{figure} [!htbp]
\includegraphics[width=0.9\columnwidth]{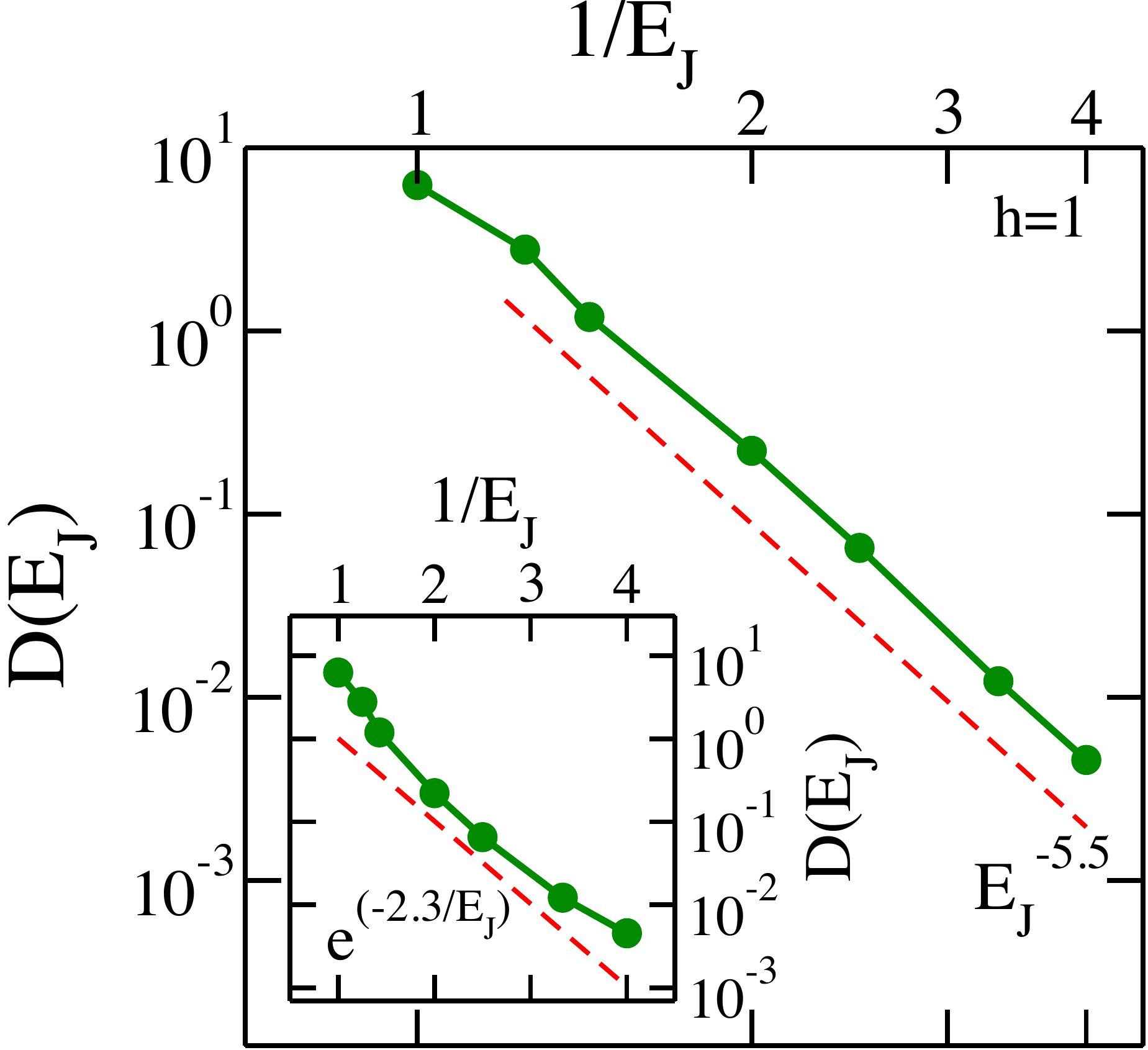}
  \caption{\label{Fig. Diffu2-n}{\footnotesize (Color online) 
 Diffusion coefficient $D$ vs $E_J$ for fixed $h=1$ in Log-Log scale (main) and Linear-Log scale (inset). 
  The red dashed lines guide the eye and represent a power-law trend $D \sim E_J^{-5.5}$ (main) 
  and an exponential trend  $D \sim \exp(-2.3/E_J)$ (inset). 
  }}
     \end{figure}

To confirm that we observe action diffusion, we rescale $\mu_2 \rightarrow \mu_2 N$ and $T \rightarrow T/N^2$ - as shown in Fig.\ref{Fig. Diffu_all-n}. 
For a given value of $E_J$ we observe very good collapse of all curves onto one master curve for 
$T \geq T_E$. 
The master curves in Fig.\ref{Fig. Diffu_all-n} 
show the turnover from diffusive $1/\sqrt{T}$ to asymptotic $1/T$ decay at the time $T_D$. The diffusion process assumes that a diffusion coefficient $D \sim N^2/T_D$ can be read
off a fit of $T_D$. 
We found the inverse $D^{-1}$ from the intersection of the fit of the  diffusive $1/\sqrt{T}$ and the asymptotic $1/T$ trends 
(marked with green dots in Fig~\ref{Fig. Diffu_all-n}). 
The measured values of the diffusion coefficient $D$ are then reported as a function of $E_J$ in Fig.\ref{Fig. Diffu2-n},
which appears to be reasonably close to a power-law over the analyzed interval. 
To check whether the asymptotic behavior of $D$ may turn into an exponential behavior rather than power law \cite{Giardna2000finite,oganesyan2009energy,Flach2011thermal,Li2015temperature,Iubini:2016coupled}  we plot the data in Linear-Log scale in the inset of
Fig.\ref{Fig. Diffu2-n}.  From the available data we conclude that a power law is more close to the obtained data.  

Our analysis shows that the dynamics of resonances starts with a diffusion process between chaotic triplets, i.e. on a
length scale $l_r \sim \sqrt{D T_E}$. After that the  diffusion continues until all fluctuations 
stored in $N/l_r$ nonresonant patches each of the size $l_r$, were exchanged and reached a given location in the system. This happens for $T_D\sim N^2/D$ \cite{Edwards_1972}.


If the above scenario is correct, we can expect to delay the diffusion, relaxation, and ergodization process, if we manage to
efficiently destroy resonances, before they had time to diffuse. To check this prediction, we take a system with $N=512$ and $E_j=0.7$ at $h=1$.
Every time interval $T_{\Lambda} \approx 10$ we randomly pick a site $n$, and increment or decrement its momentum $p_n$
by a given value $\Delta_p$ with equal probability 
\footnote{Each time one momentum is changed by a given value $\Delta_p$, the other momenta are corrected such that the total momentum $P$ is again zero. After that all momenta are slightly rescaled to reach the original energy density $h$. Hence, the distance from the integrable limit is not changed by the kick procedure}. 
On average a given site $n_0$ is reached on a time $T_{\Lambda} N \approx 5000$.
If $\Delta_p \approx E_J/h$ we expect to efficiently detune and destroy a nonlinear resonance. 
\begin{figure} [!htbp] 
            \includegraphics[scale=1.0,width=0.48\textwidth]{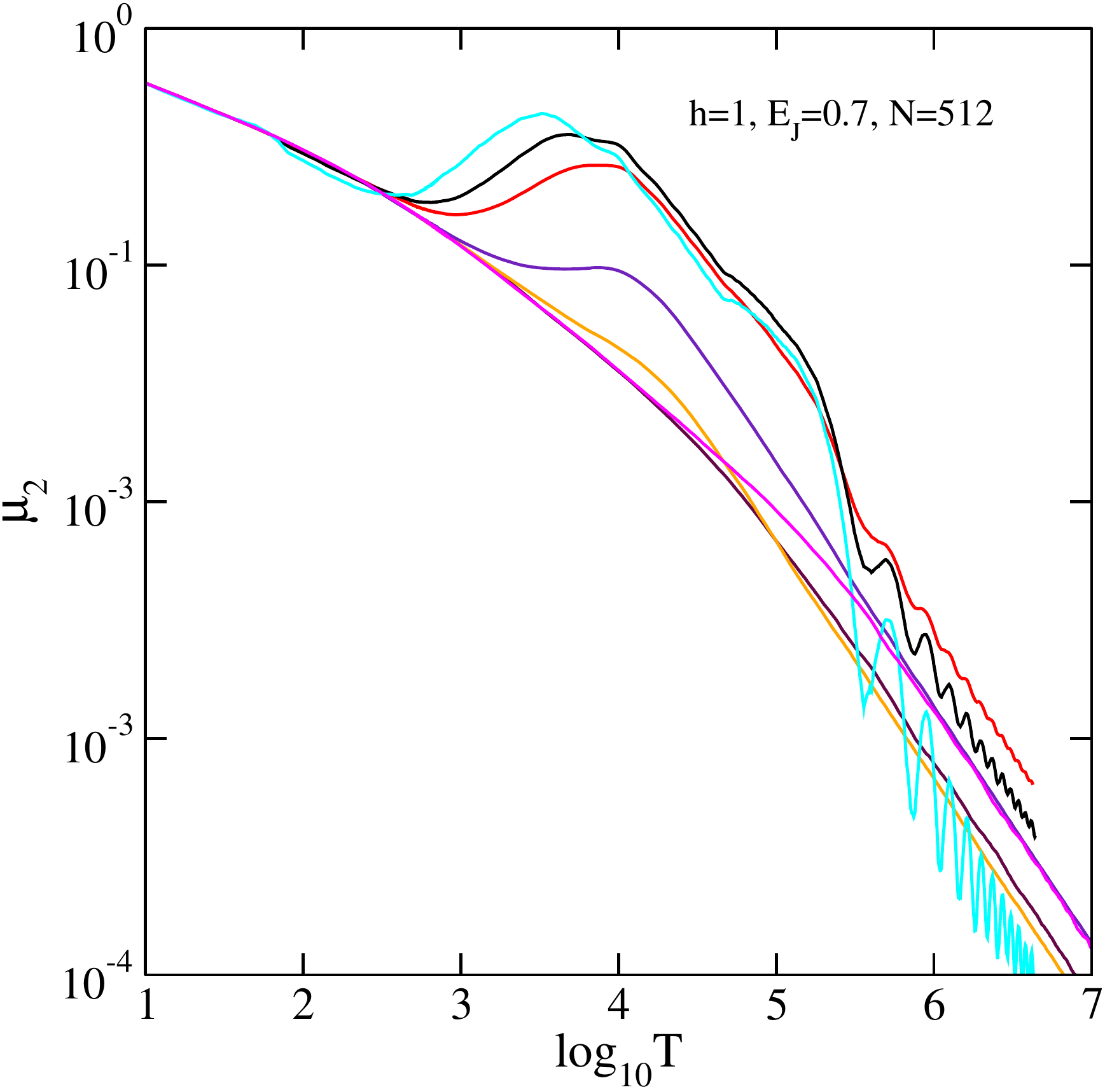}\\
  \caption{\label{Fig. kicks}{\footnotesize (Color online) The time dependence of $\mu_2(T)$ in the presence of a random kick process 
with $\Delta_p = 0$ (magenta), $0.3$ (maroon), $0.6$ (orange), $1$ (indaco), $2$ (red), $3$ (black), $5$ (cyan). 
Here $N=512$, $h=1$ and $E_J=0.7$.
  }}
     \end{figure}
The effect should become visible for $T\approx 5000$. The results in Fig.\ref{Fig. kicks} are excellently reproducing the 
prediction. The kicks will also generate new resonances at another location, such that the average number of resonances
will not change. Our results confirm that resonance diffusion is at the origin of the ergodization process. It is exactly this
diffusive process which is efficiently harmed, destroyed and delayed by the above kicking procedure.

Combining our results with previous studies shows that weakly nonintegrable many-body systems can restore ergodicity
in distinctive ways depending on the range
of the interaction network in action space. It all starts with action resonances seeding chaotic dynamics into the networks.
While long range networks provide well connected resonances with ergodization controlled
by the characteristic individual resonance chaos time scales,
short range networks instead yield a dramatic slowing down of ergodization in action space,
and lead to rare resonance diffusion. We used Josephson junction chains as a paradigmatic study case and exploited finite time average distributions to characterize the thermalizing dynamics of actions. The slowing down of the thermalization dynamics upon approaching the integrable limit results in a decreasing of an effective diffusion constant which
is related to heat conductivity. This slowing down appears to follow a power law in the distance from the integrable limit, rather than an exponential one.
We identify a novel action resonance diffusion regime responsible for the slowing down. 
The observed fragile diffusion is relying on weakly chaotic dynamics in spatially isolated action resonances. We were able 
successfully delay and suppress it
by adding weak action noise, as a proof of concept.
Among a number of intriguing open questions, we mention the search for further distinct classes of nonintegrable action networks (neither short nor long ranged), and
the impact of quantization on the fragile short range network dynamics in the vicintiy of an integrable limit.

The authors are thankful to Ara Go and Hyeong Jun Lee for assistance on computational aspects of the work. The authors acknowledge financial support from IBS (Project Code No. IBS-R024-D1). M.V.F. acknowledges the partial financial support in the framework of Increase Competitiveness Program of NUST "MISiS" K2-2020-001.

\let\itshape\upshape
\normalem
\bibliography{reference1,sergejflach}

\providecommand{\noopsort}[1]{}\providecommand{\singleletter}[1]{#1}%
\begin{thebibliography}{29}%
\makeatletter
\providecommand \@ifxundefined [1]{%
 \@ifx{#1\undefined}
}%
\providecommand \@ifnum [1]{%
 \ifnum #1\expandafter \@firstoftwo
 \else \expandafter \@secondoftwo
 \fi
}%
\providecommand \@ifx [1]{%
 \ifx #1\expandafter \@firstoftwo
 \else \expandafter \@secondoftwo
 \fi
}%
\providecommand \natexlab [1]{#1}%
\providecommand \enquote  [1]{``#1''}%
\providecommand \bibnamefont  [1]{#1}%
\providecommand \bibfnamefont [1]{#1}%
\providecommand \citenamefont [1]{#1}%
\providecommand \href@noop [0]{\@secondoftwo}%
\providecommand \href [0]{\begingroup \@sanitize@url \@href}%
\providecommand \@href[1]{\@@startlink{#1}\@@href}%
\providecommand \@@href[1]{\endgroup#1\@@endlink}%
\providecommand \@sanitize@url [0]{\catcode `\\12\catcode `\$12\catcode
  `\&12\catcode `\#12\catcode `\^12\catcode `\_12\catcode `\%12\relax}%
\providecommand \@@startlink[1]{}%
\providecommand \@@endlink[0]{}%
\providecommand \url  [0]{\begingroup\@sanitize@url \@url }%
\providecommand \@url [1]{\endgroup\@href {#1}{\urlprefix }}%
\providecommand \urlprefix  [0]{URL }%
\providecommand \Eprint [0]{\href }%
\providecommand \doibase [0]{http://dx.doi.org/}%
\providecommand \selectlanguage [0]{\@gobble}%
\providecommand \bibinfo  [0]{\@secondoftwo}%
\providecommand \bibfield  [0]{\@secondoftwo}%
\providecommand \translation [1]{[#1]}%
\providecommand \BibitemOpen [0]{}%
\providecommand \bibitemStop [0]{}%
\providecommand \bibitemNoStop [0]{.\EOS\space}%
\providecommand \EOS [0]{\spacefactor3000\relax}%
\providecommand \BibitemShut  [1]{\csname bibitem#1\endcsname}%
\let\auto@bib@innerbib\@empty
\bibitem [{\citenamefont {Huang}(1987)}]{book:huang87}%
  \BibitemOpen
  \bibfield  {author} {\bibinfo {author} {\bibfnamefont {K.}~\bibnamefont
  {Huang}},\ }\href {https://books.google.co.kr/books?id=M8PvAAAAMAAJ} {\emph
  {\bibinfo {title} {Statistical mechanics}}}\ (\bibinfo  {publisher} {Wiley},\
  \bibinfo {year} {1987})\BibitemShut {NoStop}%
\bibitem [{\citenamefont {Ford}(1992)}]{FORD1992FPU}%
  \BibitemOpen
  \bibfield  {author} {\bibinfo {author} {\bibfnamefont {Joseph}\ \bibnamefont
  {Ford}},\ }\bibfield  {title} {\enquote {\bibinfo {title} {The
  fermi-pasta-ulam problem: Paradox turns discovery},}\ }\href {\doibase
  https://doi.org/10.1016/0370-1573(92)90116-H} {\bibfield  {journal} {\bibinfo
   {journal} {Physics Reports}\ }\textbf {\bibinfo {volume} {213}},\ \bibinfo
  {pages} {271 -- 310} (\bibinfo {year} {1992})}\BibitemShut {NoStop}%
\bibitem [{\citenamefont {Weissert}(1997)}]{fpugen}%
  \BibitemOpen
  \bibfield  {author} {\bibinfo {author} {\bibfnamefont {Thomas~P}\
  \bibnamefont {Weissert}},\ }\href@noop {} {\emph {\bibinfo {title} {{The
  Genesis of Simulation in Dynamics: Pursuing the Fermi-Pasta-Ulam Problem}}}}\
  (\bibinfo  {publisher} {Springer-Verlag},\ \bibinfo {address} {New York,
  NY},\ \bibinfo {year} {1997})\BibitemShut {NoStop}%
\bibitem [{\citenamefont {Flach}\ \emph {et~al.}(2005)\citenamefont {Flach},
  \citenamefont {Ivanchenko},\ and\ \citenamefont
  {Kanakov}}]{FlachIvanchenkoKanakovPRL2005}%
  \BibitemOpen
  \bibfield  {author} {\bibinfo {author} {\bibfnamefont {S.}~\bibnamefont
  {Flach}}, \bibinfo {author} {\bibfnamefont {M.~V.}\ \bibnamefont
  {Ivanchenko}}, \ and\ \bibinfo {author} {\bibfnamefont {O.~I.}\ \bibnamefont
  {Kanakov}},\ }\bibfield  {title} {\enquote {\bibinfo {title} {$q$-breathers
  and the fermi-pasta-ulam problem},}\ }\href {\doibase
  10.1103/PhysRevLett.95.064102} {\bibfield  {journal} {\bibinfo  {journal}
  {Phys. Rev. Lett.}\ }\textbf {\bibinfo {volume} {95}},\ \bibinfo {pages}
  {064102} (\bibinfo {year} {2005})}\BibitemShut {NoStop}%
\bibitem [{\citenamefont {Gallavotti}(2007)}]{gallavotti2007}%
  \BibitemOpen
  \bibfield  {author} {\bibinfo {author} {\bibfnamefont {Giovanni}\
  \bibnamefont {Gallavotti}},\ }\href@noop {} {\emph {\bibinfo {title} {The
  Fermi-Pasta-Ulam problem: a status report}}},\ Vol.\ \bibinfo {volume} {728}\
  (\bibinfo  {publisher} {Springer},\ \bibinfo {year} {2007})\BibitemShut
  {NoStop}%
\bibitem [{\citenamefont {Danieli}\ \emph {et~al.}(2017)\citenamefont
  {Danieli}, \citenamefont {Campbell},\ and\ \citenamefont
  {Flach}}]{Danieli:2017}%
  \BibitemOpen
  \bibfield  {author} {\bibinfo {author} {\bibfnamefont {C.}~\bibnamefont
  {Danieli}}, \bibinfo {author} {\bibfnamefont {D.~K.}\ \bibnamefont
  {Campbell}}, \ and\ \bibinfo {author} {\bibfnamefont {S.}~\bibnamefont
  {Flach}},\ }\bibfield  {title} {\enquote {\bibinfo {title} {Intermittent
  many-body dynamics at equilibrium},}\ }\href {\doibase
  10.1103/PhysRevE.95.060202} {\bibfield  {journal} {\bibinfo  {journal} {Phys.
  Rev. E}\ }\textbf {\bibinfo {volume} {95}},\ \bibinfo {pages} {060202}
  (\bibinfo {year} {2017})}\BibitemShut {NoStop}%
\bibitem [{\citenamefont {Rasmussen}\ \emph {et~al.}(2000)\citenamefont
  {Rasmussen}, \citenamefont {Cretegny}, \citenamefont {Kevrekidis},\ and\
  \citenamefont {Gr\o{}nbech-Jensen}}]{rasmussen00}%
  \BibitemOpen
  \bibfield  {author} {\bibinfo {author} {\bibfnamefont {K.~\O{}.}\
  \bibnamefont {Rasmussen}}, \bibinfo {author} {\bibfnamefont {T.}~\bibnamefont
  {Cretegny}}, \bibinfo {author} {\bibfnamefont {P.~G.}\ \bibnamefont
  {Kevrekidis}}, \ and\ \bibinfo {author} {\bibfnamefont {Niels}\ \bibnamefont
  {Gr\o{}nbech-Jensen}},\ }\bibfield  {title} {\enquote {\bibinfo {title}
  {Statistical mechanics of a discrete nonlinear system},}\ }\href {\doibase
  10.1103/PhysRevLett.84.3740} {\bibfield  {journal} {\bibinfo  {journal}
  {Phys. Rev. Lett.}\ }\textbf {\bibinfo {volume} {84}},\ \bibinfo {pages}
  {3740--3743} (\bibinfo {year} {2000})}\BibitemShut {NoStop}%
\bibitem [{\citenamefont {Iubini}\ \emph {et~al.}(2013)\citenamefont {Iubini},
  \citenamefont {Franzosi}, \citenamefont {Livi}, \citenamefont {Oppo},\ and\
  \citenamefont {Politi}}]{Iubini:2013}%
  \BibitemOpen
  \bibfield  {author} {\bibinfo {author} {\bibfnamefont {S}~\bibnamefont
  {Iubini}}, \bibinfo {author} {\bibfnamefont {R}~\bibnamefont {Franzosi}},
  \bibinfo {author} {\bibfnamefont {R}~\bibnamefont {Livi}}, \bibinfo {author}
  {\bibfnamefont {G-L}\ \bibnamefont {Oppo}}, \ and\ \bibinfo {author}
  {\bibfnamefont {A}~\bibnamefont {Politi}},\ }\bibfield  {title} {\enquote
  {\bibinfo {title} {Discrete breathers and negative-temperature states},}\
  }\href {http://stacks.iop.org/1367-2630/15/i=2/a=023032} {\bibfield
  {journal} {\bibinfo  {journal} {New J. Phys.}\ }\textbf {\bibinfo {volume}
  {15}},\ \bibinfo {pages} {023032} (\bibinfo {year} {2013})}\BibitemShut
  {NoStop}%
\bibitem [{\citenamefont {Mithun}\ \emph {et~al.}(2018)\citenamefont {Mithun},
  \citenamefont {Kati}, \citenamefont {Danieli},\ and\ \citenamefont
  {Flach}}]{Mithun:2018}%
  \BibitemOpen
  \bibfield  {author} {\bibinfo {author} {\bibfnamefont {Thudiyangal}\
  \bibnamefont {Mithun}}, \bibinfo {author} {\bibfnamefont {Yagmur}\
  \bibnamefont {Kati}}, \bibinfo {author} {\bibfnamefont {Carlo}\ \bibnamefont
  {Danieli}}, \ and\ \bibinfo {author} {\bibfnamefont {Sergej}\ \bibnamefont
  {Flach}},\ }\bibfield  {title} {\enquote {\bibinfo {title} {Weakly nonergodic
  dynamics in the gross-pitaevskii lattice},}\ }\href {\doibase
  10.1103/PhysRevLett.120.184101} {\bibfield  {journal} {\bibinfo  {journal}
  {Phys. Rev. Lett.}\ }\textbf {\bibinfo {volume} {120}},\ \bibinfo {pages}
  {184101} (\bibinfo {year} {2018})}\BibitemShut {NoStop}%
\bibitem [{\citenamefont {Cherny}\ \emph {et~al.}(2019)\citenamefont {Cherny},
  \citenamefont {Engl},\ and\ \citenamefont {Flach}}]{ChernyEnglFlach19}%
  \BibitemOpen
  \bibfield  {author} {\bibinfo {author} {\bibfnamefont {Alexander~Yu.}\
  \bibnamefont {Cherny}}, \bibinfo {author} {\bibfnamefont {Thomas}\
  \bibnamefont {Engl}}, \ and\ \bibinfo {author} {\bibfnamefont {Sergej}\
  \bibnamefont {Flach}},\ }\bibfield  {title} {\enquote {\bibinfo {title}
  {Non-gibbs states on a bose-hubbard lattice},}\ }\href {\doibase
  10.1103/PhysRevA.99.023603} {\bibfield  {journal} {\bibinfo  {journal} {Phys.
  Rev. A}\ }\textbf {\bibinfo {volume} {99}},\ \bibinfo {pages} {023603}
  (\bibinfo {year} {2019})}\BibitemShut {NoStop}%
\bibitem [{\citenamefont {Iubini}\ \emph {et~al.}(2019)\citenamefont {Iubini},
  \citenamefont {Chirondojan}, \citenamefont {Oppo}, \citenamefont {Politi},\
  and\ \citenamefont {Politi}}]{Iubini:2019dynamical}%
  \BibitemOpen
  \bibfield  {author} {\bibinfo {author} {\bibfnamefont {Stefano}\ \bibnamefont
  {Iubini}}, \bibinfo {author} {\bibfnamefont {Liviu}\ \bibnamefont
  {Chirondojan}}, \bibinfo {author} {\bibfnamefont {Gian-Luca}\ \bibnamefont
  {Oppo}}, \bibinfo {author} {\bibfnamefont {Antonio}\ \bibnamefont {Politi}},
  \ and\ \bibinfo {author} {\bibfnamefont {Paolo}\ \bibnamefont {Politi}},\
  }\bibfield  {title} {\enquote {\bibinfo {title} {Dynamical freezing of
  relaxation to equilibrium},}\ }\href {\doibase
  10.1103/PhysRevLett.122.084102} {\bibfield  {journal} {\bibinfo  {journal}
  {Phys. Rev. Lett.}\ }\textbf {\bibinfo {volume} {122}},\ \bibinfo {pages}
  {084102} (\bibinfo {year} {2019})}\BibitemShut {NoStop}%
\bibitem [{\citenamefont {Moser}(2001)}]{Moser2001}%
  \BibitemOpen
  \bibfield  {author} {\bibinfo {author} {\bibfnamefont {Jurgen}\ \bibnamefont
  {Moser}},\ }\href@noop {} {\emph {\bibinfo {title} {Stable and random motions
  in dynamical systems: With special emphasis on celestial mechanics}}},\
  Vol.~\bibinfo {volume} {1}\ (\bibinfo  {publisher} {Princeton university
  press},\ \bibinfo {year} {2001})\BibitemShut {NoStop}%
\bibitem [{\citenamefont {Diacu}\ and\ \citenamefont
  {Holmes}(1999)}]{DiacuHolmes1999}%
  \BibitemOpen
  \bibfield  {author} {\bibinfo {author} {\bibfnamefont {Florin}\ \bibnamefont
  {Diacu}}\ and\ \bibinfo {author} {\bibfnamefont {Philip}\ \bibnamefont
  {Holmes}},\ }\href@noop {} {\emph {\bibinfo {title} {Celestial encounters:
  the origins of chaos and stability}}},\ Vol.~\bibinfo {volume} {22}\
  (\bibinfo  {publisher} {Princeton University Press},\ \bibinfo {year}
  {1999})\BibitemShut {NoStop}%
\bibitem [{\citenamefont {Basko}\ \emph {et~al.}(2006)\citenamefont {Basko},
  \citenamefont {Aleiner},\ and\ \citenamefont {Altshuler}}]{Basko2006metal}%
  \BibitemOpen
  \bibfield  {author} {\bibinfo {author} {\bibfnamefont {D.M.}\ \bibnamefont
  {Basko}}, \bibinfo {author} {\bibfnamefont {I.L.}\ \bibnamefont {Aleiner}}, \
  and\ \bibinfo {author} {\bibfnamefont {B.L.}\ \bibnamefont {Altshuler}},\
  }\bibfield  {title} {\enquote {\bibinfo {title} {Metal insulator transition
  in a weakly interacting many-electron system with localized single-particle
  states},}\ }\href {\doibase https://doi.org/10.1016/j.aop.2005.11.014}
  {\bibfield  {journal} {\bibinfo  {journal} {Annals of Physics}\ }\textbf
  {\bibinfo {volume} {321}},\ \bibinfo {pages} {1126 -- 1205} (\bibinfo {year}
  {2006})}\BibitemShut {NoStop}%
\bibitem [{\citenamefont {Nandkishore}\ and\ \citenamefont
  {Huse}(2015)}]{NandkishoreHuse2015}%
  \BibitemOpen
  \bibfield  {author} {\bibinfo {author} {\bibfnamefont {Rahul}\ \bibnamefont
  {Nandkishore}}\ and\ \bibinfo {author} {\bibfnamefont {David~A}\ \bibnamefont
  {Huse}},\ }\bibfield  {title} {\enquote {\bibinfo {title} {Many-body
  localization and thermalization in quantum statistical mechanics},}\
  }\href@noop {} {\bibfield  {journal} {\bibinfo  {journal} {Annu. Rev.
  Condens. Matter Phys.}\ }\textbf {\bibinfo {volume} {6}},\ \bibinfo {pages}
  {15--38} (\bibinfo {year} {2015})}\BibitemShut {NoStop}%
\bibitem [{\citenamefont {Escande}\ \emph {et~al.}(1994)\citenamefont
  {Escande}, \citenamefont {Kantz}, \citenamefont {Livi},\ and\ \citenamefont
  {Ruffo}}]{Escande:1994}%
  \BibitemOpen
  \bibfield  {author} {\bibinfo {author} {\bibfnamefont {Dominique}\
  \bibnamefont {Escande}}, \bibinfo {author} {\bibfnamefont {Holger}\
  \bibnamefont {Kantz}}, \bibinfo {author} {\bibfnamefont {Roberto}\
  \bibnamefont {Livi}}, \ and\ \bibinfo {author} {\bibfnamefont {Stefano}\
  \bibnamefont {Ruffo}},\ }\bibfield  {title} {\enquote {\bibinfo {title}
  {Self-consistent check of the validity of gibbs calculus using dynamical
  variables},}\ }\href {\doibase 10.1007/BF02188677} {\bibfield  {journal}
  {\bibinfo  {journal} {Journal of Statistical Physics}\ }\textbf {\bibinfo
  {volume} {76}},\ \bibinfo {pages} {605--626} (\bibinfo {year}
  {1994})}\BibitemShut {NoStop}%
\bibitem [{\citenamefont {Pino}\ \emph {et~al.}(2016)\citenamefont {Pino},
  \citenamefont {Ioffe},\ and\ \citenamefont {Altshuler}}]{Pino:2016}%
  \BibitemOpen
  \bibfield  {author} {\bibinfo {author} {\bibfnamefont {Manuel}\ \bibnamefont
  {Pino}}, \bibinfo {author} {\bibfnamefont {Lev~B.}\ \bibnamefont {Ioffe}}, \
  and\ \bibinfo {author} {\bibfnamefont {Boris~L.}\ \bibnamefont {Altshuler}},\
  }\bibfield  {title} {\enquote {\bibinfo {title} {Nonergodic metallic and
  insulating phases of josephson junction chains},}\ }\href {\doibase
  10.1073/pnas.1520033113} {\bibfield  {journal} {\bibinfo  {journal} {PNAS}\
  }\textbf {\bibinfo {volume} {113}},\ \bibinfo {pages} {536--541} (\bibinfo
  {year} {2016})}\BibitemShut {NoStop}%
\bibitem [{\citenamefont {Mithun}\ \emph {et~al.}(2019)\citenamefont {Mithun},
  \citenamefont {Danieli}, \citenamefont {Kati},\ and\ \citenamefont
  {Flach}}]{Mithun:2019}%
  \BibitemOpen
  \bibfield  {author} {\bibinfo {author} {\bibfnamefont {Thudiyangal}\
  \bibnamefont {Mithun}}, \bibinfo {author} {\bibfnamefont {Carlo}\
  \bibnamefont {Danieli}}, \bibinfo {author} {\bibfnamefont {Yagmur}\
  \bibnamefont {Kati}}, \ and\ \bibinfo {author} {\bibfnamefont {Sergej}\
  \bibnamefont {Flach}},\ }\bibfield  {title} {\enquote {\bibinfo {title}
  {Dynamical glass and ergodization times in classical josephson junction
  chains},}\ }\href {\doibase 10.1103/PhysRevLett.122.054102} {\bibfield
  {journal} {\bibinfo  {journal} {Phys. Rev. Lett.}\ }\textbf {\bibinfo
  {volume} {122}},\ \bibinfo {pages} {054102} (\bibinfo {year}
  {2019})}\BibitemShut {NoStop}%
\bibitem [{\citenamefont {Danieli}\ \emph
  {et~al.}(2019{\natexlab{a}})\citenamefont {Danieli}, \citenamefont {Mithun},
  \citenamefont {Kati}, \citenamefont {Campbell},\ and\ \citenamefont
  {Flach}}]{Danieli2019dynamical}%
  \BibitemOpen
  \bibfield  {author} {\bibinfo {author} {\bibfnamefont {Carlo}\ \bibnamefont
  {Danieli}}, \bibinfo {author} {\bibfnamefont {Thudiyangal}\ \bibnamefont
  {Mithun}}, \bibinfo {author} {\bibfnamefont {Yagmur}\ \bibnamefont {Kati}},
  \bibinfo {author} {\bibfnamefont {David~K.}\ \bibnamefont {Campbell}}, \ and\
  \bibinfo {author} {\bibfnamefont {Sergej}\ \bibnamefont {Flach}},\ }\bibfield
   {title} {\enquote {\bibinfo {title} {Dynamical glass in weakly nonintegrable
  klein-gordon chains},}\ }\href {\doibase 10.1103/PhysRevE.100.032217}
  {\bibfield  {journal} {\bibinfo  {journal} {Phys. Rev. E}\ }\textbf {\bibinfo
  {volume} {100}},\ \bibinfo {pages} {032217} (\bibinfo {year}
  {2019}{\natexlab{a}})}\BibitemShut {NoStop}%
\bibitem [{\citenamefont {Chirikov}(1979)}]{Chirikov:1979}%
  \BibitemOpen
  \bibfield  {author} {\bibinfo {author} {\bibfnamefont {Boris~V}\ \bibnamefont
  {Chirikov}},\ }\bibfield  {title} {\enquote {\bibinfo {title} {A universal
  instability of many-dimensional oscillator systems},}\ }\href@noop {}
  {\bibfield  {journal} {\bibinfo  {journal} {Physics reports}\ }\textbf
  {\bibinfo {volume} {52}},\ \bibinfo {pages} {263--379} (\bibinfo {year}
  {1979})}\BibitemShut {NoStop}%
\bibitem [{\citenamefont {Danieli}\ \emph
  {et~al.}(2019{\natexlab{b}})\citenamefont {Danieli}, \citenamefont {Manda},
  \citenamefont {Mithun},\ and\ \citenamefont
  {Skokos}}]{danieli2019computational}%
  \BibitemOpen
  \bibfield  {author} {\bibinfo {author} {\bibfnamefont {C.}~\bibnamefont
  {Danieli}}, \bibinfo {author} {\bibfnamefont {B.~Many}\ \bibnamefont
  {Manda}}, \bibinfo {author} {\bibfnamefont {T.}~\bibnamefont {Mithun}}, \
  and\ \bibinfo {author} {\bibfnamefont {Ch.}\ \bibnamefont {Skokos}},\
  }\bibfield  {title} {\enquote {\bibinfo {title} {Computational efficiency of
  numerical integration methods for the tangent dynamics of many-body
  hamiltonian systems in one and two spatial dimensions},}\ }\href {\doibase
  http://dx.doi.org/10.3934/mine.2019.3.447} {\bibfield  {journal} {\bibinfo
  {journal} {Mathematics in Engineering}\ }\textbf {\bibinfo {volume} {1}},\
  \bibinfo {pages} {447} (\bibinfo {year} {2019}{\natexlab{b}})}\BibitemShut
  {NoStop}%
\bibitem [{\citenamefont {Basko}(2012)}]{Basko_2012}%
  \BibitemOpen
  \bibfield  {author} {\bibinfo {author} {\bibfnamefont {D.~M.}\ \bibnamefont
  {Basko}},\ }\bibfield  {title} {\enquote {\bibinfo {title} {Local nature and
  scaling of chaos in weakly nonlinear disordered chains},}\ }\href {\doibase
  10.1103/PhysRevE.86.036202} {\bibfield  {journal} {\bibinfo  {journal} {Phys.
  Rev. E}\ }\textbf {\bibinfo {volume} {86}},\ \bibinfo {pages} {036202}
  (\bibinfo {year} {2012})}\BibitemShut {NoStop}%
\bibitem [{\citenamefont {Giardin\`a}\ \emph {et~al.}(2000)\citenamefont
  {Giardin\`a}, \citenamefont {Livi}, \citenamefont {Politi},\ and\
  \citenamefont {Vassalli}}]{Giardna2000finite}%
  \BibitemOpen
  \bibfield  {author} {\bibinfo {author} {\bibfnamefont {C.}~\bibnamefont
  {Giardin\`a}}, \bibinfo {author} {\bibfnamefont {R.}~\bibnamefont {Livi}},
  \bibinfo {author} {\bibfnamefont {A.}~\bibnamefont {Politi}}, \ and\ \bibinfo
  {author} {\bibfnamefont {M.}~\bibnamefont {Vassalli}},\ }\bibfield  {title}
  {\enquote {\bibinfo {title} {Finite thermal conductivity in 1d lattices},}\
  }\href {\doibase 10.1103/PhysRevLett.84.2144} {\bibfield  {journal} {\bibinfo
   {journal} {Phys. Rev. Lett.}\ }\textbf {\bibinfo {volume} {84}},\ \bibinfo
  {pages} {2144--2147} (\bibinfo {year} {2000})}\BibitemShut {NoStop}%
\bibitem [{\citenamefont {Oganesyan}\ \emph {et~al.}(2009)\citenamefont
  {Oganesyan}, \citenamefont {Pal},\ and\ \citenamefont
  {Huse}}]{oganesyan2009energy}%
  \BibitemOpen
  \bibfield  {author} {\bibinfo {author} {\bibfnamefont {Vadim}\ \bibnamefont
  {Oganesyan}}, \bibinfo {author} {\bibfnamefont {Arijeet}\ \bibnamefont
  {Pal}}, \ and\ \bibinfo {author} {\bibfnamefont {David~A.}\ \bibnamefont
  {Huse}},\ }\bibfield  {title} {\enquote {\bibinfo {title} {Energy transport
  in disordered classical spin chains},}\ }\href {\doibase
  10.1103/PhysRevB.80.115104} {\bibfield  {journal} {\bibinfo  {journal} {Phys.
  Rev. B}\ }\textbf {\bibinfo {volume} {80}},\ \bibinfo {pages} {115104}
  (\bibinfo {year} {2009})}\BibitemShut {NoStop}%
\bibitem [{\citenamefont {Flach}\ \emph {et~al.}(2011)\citenamefont {Flach},
  \citenamefont {Ivanchenko}, \citenamefont {Li},\ and\ \citenamefont
  {Li}}]{Flach2011thermal}%
  \BibitemOpen
  \bibfield  {author} {\bibinfo {author} {\bibfnamefont {Sergej}\ \bibnamefont
  {Flach}}, \bibinfo {author} {\bibfnamefont {Mikhail}\ \bibnamefont
  {Ivanchenko}}, \bibinfo {author} {\bibfnamefont {Nianbei}\ \bibnamefont
  {Li}}, \ and\ \bibinfo {author} {\bibfnamefont {Baowen}\ \bibnamefont {Li}},\
  }\bibfield  {title} {\enquote {\bibinfo {title} {Thermal conductivity of
  nonlinear waves in disordered chains},}\ }\href {\doibase
  10.1007/s12043-011-0186-0} {\bibfield  {journal} {\bibinfo  {journal} {The
  European Physical Journal B}\ }\textbf {\bibinfo {volume} {77}},\ \bibinfo
  {pages} {1007} (\bibinfo {year} {2011})}\BibitemShut {NoStop}%
\bibitem [{\citenamefont {Li}\ \emph {et~al.}(2015)\citenamefont {Li},
  \citenamefont {Li},\ and\ \citenamefont {Li}}]{Li2015temperature}%
  \BibitemOpen
  \bibfield  {author} {\bibinfo {author} {\bibfnamefont {Yunyun}\ \bibnamefont
  {Li}}, \bibinfo {author} {\bibfnamefont {Nianbei}\ \bibnamefont {Li}}, \ and\
  \bibinfo {author} {\bibfnamefont {Baowen}\ \bibnamefont {Li}},\ }\bibfield
  {title} {\enquote {\bibinfo {title} {Temperature dependence of thermal
  conductivities of coupled rotator lattice and the momentum diffusion in
  standard map},}\ }\href {\doibase 10.1140/epjb/e2015-60361-5} {\bibfield
  {journal} {\bibinfo  {journal} {The European Physical Journal B}\ }\textbf
  {\bibinfo {volume} {88}},\ \bibinfo {pages} {182} (\bibinfo {year}
  {2015})}\BibitemShut {NoStop}%
\bibitem [{\citenamefont {Iubini}\ \emph {et~al.}(2016)\citenamefont {Iubini},
  \citenamefont {Lepri}, \citenamefont {Livi},\ and\ \citenamefont
  {Politi}}]{Iubini:2016coupled}%
  \BibitemOpen
  \bibfield  {author} {\bibinfo {author} {\bibfnamefont {S}~\bibnamefont
  {Iubini}}, \bibinfo {author} {\bibfnamefont {S}~\bibnamefont {Lepri}},
  \bibinfo {author} {\bibfnamefont {R}~\bibnamefont {Livi}}, \ and\ \bibinfo
  {author} {\bibfnamefont {A}~\bibnamefont {Politi}},\ }\bibfield  {title}
  {\enquote {\bibinfo {title} {Coupled transport in rotor models},}\ }\href
  {\doibase 10.1088/1367-2630/18/8/083023} {\bibfield  {journal} {\bibinfo
  {journal} {New Journal of Physics}\ }\textbf {\bibinfo {volume} {18}},\
  \bibinfo {pages} {083023} (\bibinfo {year} {2016})}\BibitemShut {NoStop}%
\bibitem [{\citenamefont {Edwards}\ and\ \citenamefont
  {Thouless}(1972)}]{Edwards_1972}%
  \BibitemOpen
  \bibfield  {author} {\bibinfo {author} {\bibfnamefont {J~T}\ \bibnamefont
  {Edwards}}\ and\ \bibinfo {author} {\bibfnamefont {D~J}\ \bibnamefont
  {Thouless}},\ }\bibfield  {title} {\enquote {\bibinfo {title} {Numerical
  studies of localization in disordered systems},}\ }\href {\doibase
  10.1088/0022-3719/5/8/007} {\bibfield  {journal} {\bibinfo  {journal}
  {Journal of Physics C: Solid State Physics}\ }\textbf {\bibinfo {volume}
  {5}},\ \bibinfo {pages} {807--820} (\bibinfo {year} {1972})}\BibitemShut
  {NoStop}%
\bibitem [{Note1()}]{Note1}%
  \BibitemOpen
  \bibinfo {note} {Each time one momentum is changed by a given value $\Delta
  _p$, the other momenta are corrected such that the total momentum $P$ is
  again zero. After that all momenta are slightly rescaled to reach the
  original energy density $h$. Hence, the distance from the integrable limit is
  not changed by the kick procedure}\BibitemShut {NoStop}%
\end{thebibliography}%

 \end{document}